\documentclass[]{spie}  
\usepackage[]{graphicx}

\title{Numerical control matrix rotation for the LINC-NIRVANA 
Multi-Conjugate Adaptive Optics system}

\author{Carmelo Arcidiacono\supit{a,b} Thomas Bertram\supit{c} Roberto Ragazzoni\supit{b}  Jacopo Farinato\supit{b} Simone Esposito\supit{a} Armando Riccardi\supit{a} Enrico Pinna\supit{a} Alfio Puglisi\supit{a} Luca Fini\supit{a}
Marco Xompero\supit{a} Lorenzo Busoni\supit{a} Fernando Quiros-Pacheco\supit{a} and Runa Briguglio\supit{a}
\skiplinehalf
\supit{a}INAF-Osservatorio Astrofisico di Arcetri, Largo Enrico Fermi 5, I-50125 Firenze, Italy; \\
\supit{b}INAF-Osservatorio Astronomico di Padova, Vicolo dell'Osservatorio 5, I-35122 Padova, Italy; \\
\supit{c} Max-Planck-Institute for Astronomy, Koenigstuhl 17, D-69117 Heidelberg, Germany \\
}

\authorinfo{Further author information: (Send correspondence to Carmelo Arcidiacono)\\C.A: E-mail: carmelo@arcetri.astro.it, Telephone: +39 055 2752 293}

\newcommand\aap{{A\&A}}%
\newcommand\pasp{{PASP}}%
\begin{document}
  \def\arcmin{$^{\prime}$}
  \def\arcsec{$^{\prime\prime}$}
  \def\arcdeg{$^{\rm o}$}
  \def\Ib{{\bf I} (0.85 $\mu${\em m})$\;$}
  \def\Jb{{\bf J} (1.25 $\mu${\em m})$\;$}
  \def\Hb{{\bf H} (1.65 $\mu${\em m})$\;$}
  \def\Kb{{\bf Ks} (2.12 $\mu${\em m})$\;$}
  \def\spie{Proc. of SPIE}
  \def\aap{A\&A}
  \def\pasp{PASP}
  \def\byone{1\arcmin$\times$1\arcmin}
  \def\bytwo{2\arcmin$\times$2\arcmin}
  \maketitle

\begin{abstract}
LINC-NIRVANA will realize the interferometric imaging focal station of the Large Binocular Telescope. A double Layer Oriented multi-conjugate adaptive optics system assists the two arms of the interferometer, supplying high order wave-front correction. In order to counterbalance the field rotation, mechanical derotation for the two ground wave-front sensors, and optical derotators for the mid-high layers sensors fix the positions of the focal planes with respect to the pyramids aboard the wave-front sensors. The derotation introduces pupil images rotation on the wavefront sensors: the projection of the deformable mirrors on the sensor consequently change. The proper adjustment of the control matrix will be applied in real-time through numerical computation of the new matrix. In this paper we investigate the temporal and computational aspects related to the pupils rotation, explicitly computing the wave-front errors that may be generated.
\end{abstract}


\keywords{Adaptive Optics, LBT, LINC-NIRVANA}
\section{Introduction}
\label{sec:intro}  
The foreseen interferometric imaging focal plane of the Large Binocular Telescope\cite{2010ApOpt..49..115H} (LBT) will be realized by the LINC-NIRVANA instrument. 	LINC-NIRVANA (The LBT INterferometric Camera and Near– IR/Visible Adaptive INterferometer for
Astronomy)\cite{2003SPIE.4839..536R,2005CRPhy...6.1129G,2008SPIE.7013E..65H} is a near-infrared Fizeau interferometer for the LBT and will combine the light from the
	two 8.4\,m primary mirrors into one single focus. The two primary mirrors are mounted on a common alt-azimuthal mount with a center-to-center separation of 14.4\,m. In order to compensate for the atmospheric 
	distortions of the incoming wavefront and to maintain a zero optical path difference (OPD) between its two arms, LINC-NIRVANA will use a Multi-Conjugate Adaptive Optics\cite{beckers88,beckers89a} (MCAO)  system\cite{2006SPIE.6272E..70F,2008SPIE.7015E.149F} implementing the Multiple Field of View\cite{mfov} version of the Layer Oriented\cite{LO1,LO2} technique and a fringe-tracker\cite{2004SPIE.5491.1454B,2006SPIE.6268E.117B}. Since the coherent combination 
	of the light happens in the focal plane, LINC-NIRVANA will have a very large, compared to Michelson interferometers, field-of-view 
	of 10\arcsec$\times$10\arcsec. In particular the MCAO system uses as wave-front sensor a multi-pyramid system covering a Field of View (FoV) of 6\arcmin. The resulting Point Spread Function (PSF) is the combination of a 8.2\,m PSF modulated by the fringes corresponding to the 14.4\,m separation of the two primary mirrors (8.2 is the corresponding primary dimension vignetted by the secondary mirror, which is the pupil). 
	The diameter of the larger Airy disk is given by $\alpha_{Airy}=1.22\lambda/D$, with the single telescope diameter $D$. In contrast the separation of the fringes depends on the separation $b$ 
	of the two primary mirrors is given by $\alpha_{fringes}=1.22\lambda/b$. The angular 
	resolution in the image thus corresponds to a 22.8\,m telescope in the projected direction connecting the two primary mirrors and of an 8.2\,m telescope in the perpendicular direction. By combining several images, each taken at a different paralactic angle and thus orientation of the fringes on the sky, the full angular resolution in all directions can then be reconstructed using deconvolution algorithms\cite{2007A&A...471.1091L}.
LINC-NIRVANA will be capable of obtaining diffraction-limited images in the wavelength range from J to K 
	band\cite{2006SPIE.6269E..11B}. With a pixel scale of 0.005\arcsec/pixel for Nyquist sampling of 
	the fringes in J-band, the field-of-view is essentially limited by the size of the detector to 10\arcsec$\times$10\arcsec.
LBT is a double adaptive telescope in which the standard static secondary mirror\cite{2003SPIE.5169..159R} is replaced by an high order adaptive secondary mirror controlled by 672 actuators.
The instrument is mounted on the telescope and it sees the field rotation typical of alt-az systems. Since the focal plane rotates with respect to the pyramids system LINC-NIRVANA provides the derotation of the field at the level of the Wave-Front Sensors (WFSs). In this way the pupil is rotated, therefore the projection of the adaptive secondary mirror changes on the WFSs.
In this paper we study the effect of this rotation on the performance and propose the best strategy to solve the problem.
\section{Adaptive Optics on LINC-NIRVANA}
\label{sec:ao}  
\begin{figure}
   \begin{center}
   \begin{tabular}{c}
   \includegraphics[width=16.5cm]{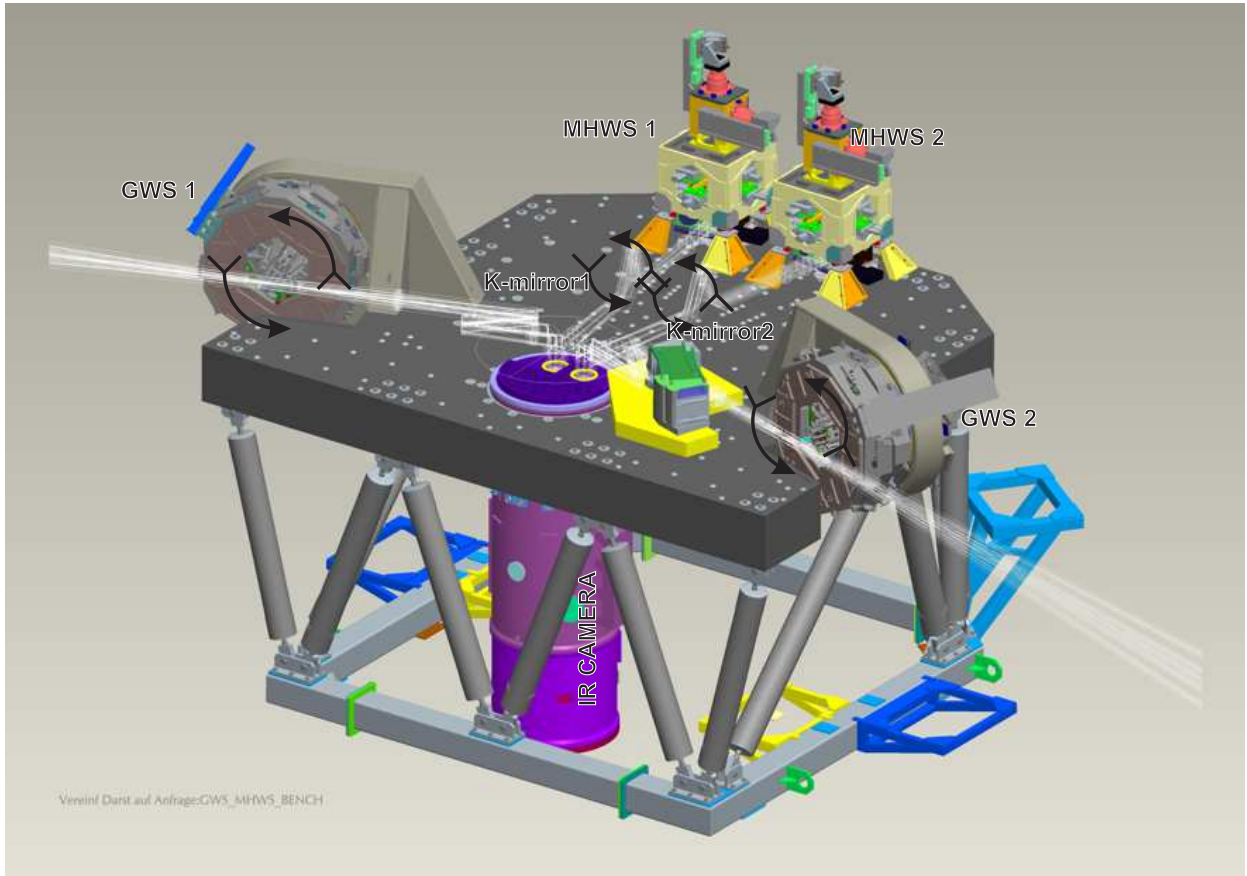}
   \end{tabular}
   \end{center}
   \caption[fig2]
   { \label{fig:fig2}
The picture shows a opto-mechanical CAD-view of the LINC-NIRVANA bench. The lines crossing the bench are the two optical beams coming from the two bent Gregorian Foci of the LBT. The four wave-front sensing devices are labeled, GWS 1 and 2 and MHWS 1 and 2. On the GWS the arrows underline the rotation movements. In the picture also the K-mirrors (1 and 2) are indicated. Also in this case the arrows represents the movement capability.}
   \end{figure}

The LINC-NIRVANA Fizeau interferometer takes advantage of the MCAO correction to enlarge the isopistonic patch size and to increase in this way the sky-coverage of the fringe tracker sensor. The MCAO system will be the first realization of the Multiple Field of View version of the Layer Oriented concept. The  latter has been successfully proven on sky by the ESO Multiconjugate Adaptive Optics Demonstrator\cite{2008SPIE.7015E.155A} (MAD) in 2007. In MAD the correction was applied by a couple of Deformable Mirrors (DM) mounted on the instruments bench (at the Nasmyth of the VLT): to correct for field rotation an optical derorator rotated the field before the DM. In this way the projection of the DM on the wavefront sensors was fixed. On LINC-NIRVANA (see Figure~\ref{fig:fig2}) two (couples of) different multi-pyramid sensors sense the turbulence being conjugated at the ground layer (Ground Wavefront Sensor, GWS) and to the high layers (MidHigh Wavefront Sensor, MHWS). While GWS mechanically rotates to counterbalance field rotation the MHWS is mounted after a K-mirror optical derotator. In this way the focal plane projection of the field is kept fixed with respect to the pyramids. The side effect of those movements is the rotation of the pupil image on the wavefront sensors: the projection of the actuators controlling the deformable mirrors shape rotates on the pupil plane projected on the wavefront sensing CCD (both the adaptive secondary mirror and the deformable mirror on the bench), loosing in this way the coupling between sub-apertures and actuators. 
Here we produced an error budget-like analysis of this effect in order to identify the best strategy for correcting it.


\section{The measurements}
\label{sec:meas}
\begin{figure}
   \begin{center}
   \begin{tabular}{c}
   \includegraphics[width=14cm]{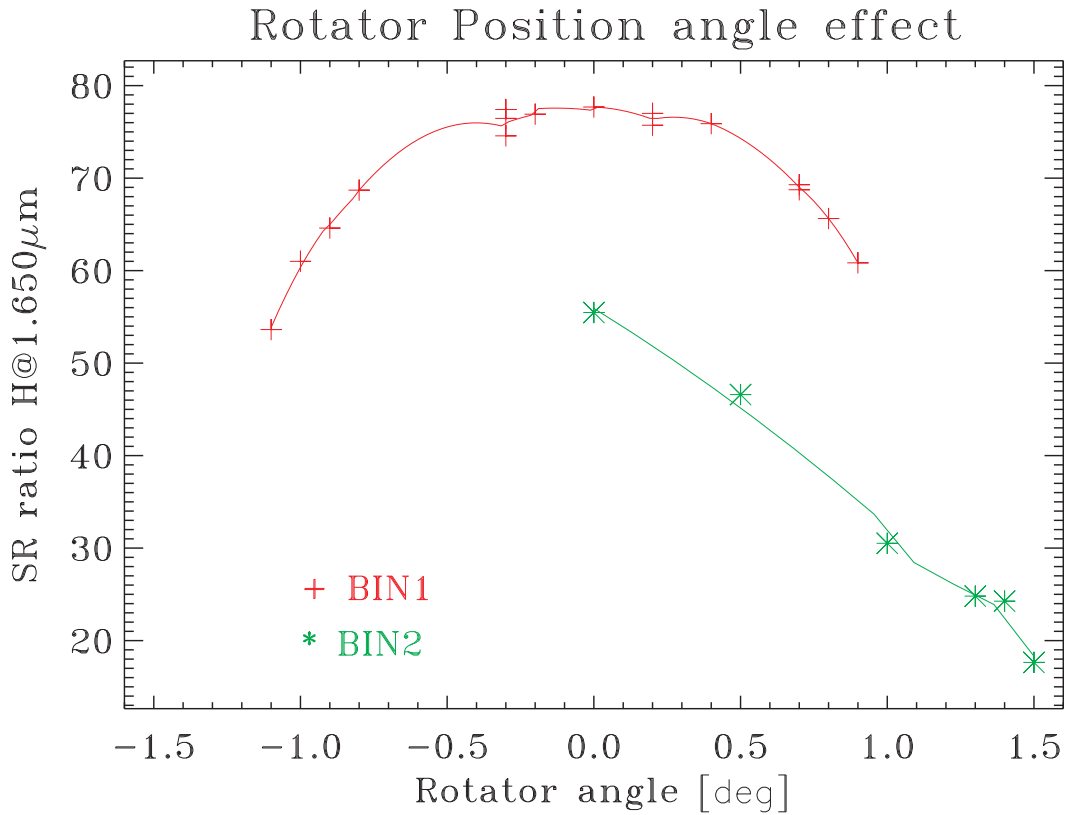}
   \end{tabular}
   \end{center}
   \caption[fig1]
   { \label{fig:fig1}
The figure shows the measured relation between the derotation angle and the performance for a given turbulence of 0.8\arcsec seeing in V band and a wind speed of 15m/sec.}
   \end{figure}

The First Light Adaptive Optics\cite{FLAO} (FLAO) system for the LBT is using a pyramid wave-front sensor and the Adaptive Secondary (AdSec) mirror\cite{2008SPIE.7015E..27R}. In a certain sense the configuration used is similar to the Ground Layer wavefront sensor system of LINC-NIRVANA. The differences between the two systems stay mainly in the higher spatial resolution on the pupil sampling of the FLAO (30$\times$30 against 24$\times$24 subapertures) and the number of pyramids (1~vs~12), but what is important here is the different approach to the pupil rotation. The sensing system of the FLAO is mounted on a board in the Acquisition, Guiding and Wave-front (AGW) sensing unit\cite{AGW} on one of the bent-Gregorian foci of LBT. The AGW optically rerotates the pupil on the wave-front sensor using a small copy of the K-mirror mounted on the LINC-NIRVANA bench. In the FLAO case the pupil is kept fixed with respect to the CCD array.
The FLAO was recently mounted on the LBT telescope and performed successfully on sky operations: the calibration of the system is performed day-time using a retroreflector which allows the pupil reimaging on both the AdSec and on the pyramid wave-front sensor. We allocated time to measure the effect of the pupil rotation in closed loop using such as configuration.
The AdSec was both generating the turbulence (applying a moving phase) with a corresponding seeing of 0.8\arcsec and wind speed of 15\,m/sec and applying the correction using 400 and 153 Karhunen-Lo\`eve (KL) modes respectively in binning 1 (30$\times$30) and 2 (15$\times$15). We measured on the InfraRed Test Camera\cite{irtc} (IRTC) the closed loop Strehl Ratio H-band values as a function of the derotation error applied by clocking the K-mirror on the board.
We preferred to use a lab-like setup with respect to perform measurements on sky in order to have similar atmospheric condition for all measurements and in order to avoid extra-derotation errors which may be introduced by the wobble and run-out of big rotator flange that mounts the AGW.
While a full rotation of the flange corresponds to a full rotation of the pupil, for the K-mirror rotator in the wave-front sensing board of AGW this corresponds to two. 
We measured the effect for two different binnings of the wave-front sensing CCD. In fact we expect that low order correction, using less KL modes and larger spatial sampling, is more robust than the higher one, being the rotation effect visible at the edge of the pupil affecting the spatial scales domain corresponding to the shift and to structure of the modes smaller than this. Because of time scheduling we had the opportunity to fully test the binning~1 case and to perform only stability test on binning~2.
To be noticed that interaction matrix for the FLAO is measured in a position which is optimizing the actuators visibility by the wave-front sensor: because of the different sampling and geometry of the mirror actuators and WFS sub-apertures, respectively on the AdSec and on the wave-front sensing CCD, the mapping is not perfect. Therefore we selected the best pupil position to match as many actuators as possible on the sub-apertures grid, optimizing the number of controllable (visible) modes and reducing the waffle (invisible) ones. This operation is particularly important at Binning 1 and less for decreasing spatial samplings. Results are shown in Figure~\ref{fig:fig1}.

\section{Results and Conclusions}
The measurements performed at two different binnings of the CCD (Bin1 30$\times$30 sub-ap 400 modes applied, Bin2 15$\times$15 sub-ap, 153 modes applied) show that loop is almost insensitive to angle shifts (of the K-mirror) up to +/-0.3~degrees and can stay closed up to +/-1~degree far from the position on which the interaction matrix has been measured. The maximum speed which the rotators moves (close to Zenith angle) is of the order of 0.05deg/sec. This speed is quite slow and allows to up-load to the  AdSec BCU the control matrix which properly takes into account the new angle (the new projection of the pupil on the wavefront sensor).
To be underlined that on the BCU on board the AdSec available memory space is enough to save two different control matrices, the one to be instantaneously applied and the one to be used after.

The next step is to evaluate the optimal computational method of the control matrix to be up-load and evaluate the range of applicability of matrix interpolation for this purposes. In fact because of the limited spatial sampling is not always true that all the actuators are seen by the wavefront sensor, therefore since rotation is changing their visibility, the numerical interpolation is not straightforward.

\acknowledgments     
This paper is the results of the collaboration of the Italian, German and US institutes involved in the LBT project. In particular LINC-NIRVANA is learning the lesson taught by the AGW experience with both pyramids and adaptive secondary mirrors. For this reason we would like to thank to the LINC-NIRVANA and AGW teams and in particular J. Little and the LBT crew, which fully supported us for mounting and dismounting the retroreflector in front of the secondary mirror (among other tasks).


\bibliography{spie}   
\bibliographystyle{spiebib}   

\end{document}